\RequirePackage{fix-cm}
\documentclass[smallextended]{svjour3}       
\smartqed  
\usepackage{graphicx}
\begin{document}

\title{First results with the boloSource() algorithm: Photometry of faint standard stars observed by Herschel/PACS 
}


\author{MARTON, G.         \and
        Vavrek, R. \and
        Kiss, Cs. \and
        M\"uller, Th. G. 
}


\institute{G. Marton, Cs. Kiss \at
              Konkoly Observatory, Research Center for Astronomy and Earth Sciences, Hungarian Academy of Sciences; Konkoly Thege 15-17, 1121 Budapest, Hungary\\
              \email{marton.gabor@csfk.mta.hu}           
           \and
           R. Vavrek \at
              ESA/ESAC PO Box 50727 Villafranca del Castillo, 28080 Madrid, Spain
           \and
            Th. G. M\"uller \at
            Max-Planck-Institut f\"ur extraterrestrische Physik (MPE), Giessenbachstrasse, 85748, Garching, Germany 
}

\date{Received: date / Accepted: date}

\maketitle

\begin{abstract}
The boloSource() algorithm is a tool to separate the signal of compact sources from that of the diffuse background in the timeline of far-infrared measurements performed by the PACS camera of the Herschel Space Observatory. An important characteristic and quality indicator of this method is that how well it can reproduce the flux of faint standard stars which have reliable flux estimates. For this propose we selected a few calibrator targets and constructed light curves by extracting point source flux for each repetition of the measurements independently using standard aperture photometry methods. These were compared with the light curves obtained using the boloSource() method on the same dataset. The results indicate that boloSource() provides a similar level of photometric accuracy and reproducibility as the usual flux extraction and photometry methods. This new technique will be developed further and also tested against other methods in more complex fields with the goal to make it usable for large-scale studies in the future.

\end{abstract}

\section{Introduction}
\label{intro}
The original goal of the development of the boloSource() algorithm  (Vavrek et al., in prep.) is to  create a tool which can provide source-free (i.e. background only) maps of the Herschel Space Observatory observations. This kind of maps are useful because analysis of the extended emission requires maps that are free of contaminating sources. Even in the Fourier space, compact objects (the extension of the source is around, or smaller than the beam FWHM at given wavelength) contribute to the image power spectra with a significant power at a broad range of spatial frequencies, and they modify the image at frequencies comparable to the beam-size. Depending on the surface density and the clustering strength, lower spatial frequencies are contaminated with a smaller power density but typically at large bandwidth. Detailed description of power spectra calculated from Herschel observations can be found in \cite{hps}. Image analysis techniques are difficult to compare if sources are not subtracted, because their sensitivity to discrete sub-structures may be quite different. Techniques using sparsity information could be falsified by even a few point sources, while techniques analyzing full intensity maps are more sensitive to clustering. For extended emission analysis we need a technique to subtract sources that fall within a well defined range of spatial frequencies and fulfill a major requirement: preserve noise properties of the image! The classical way to do that is to try to model the source intensity $I(x,y)$ in the Level 2 (L2, \cite{hom}) position-position space. But it is possible to reduce the problem to 1D in the Level 1 (L1, \cite{hom}) detector timeline. The sources can be subtracted from the detector timeline, followed by a re-projection of the source-free image.


\section{Concept \& method}
\label{sec:1}

\subsection{Timeline interpolation}
The detailed description of the algorithm is a subject of a future paper (Vavrek et al., in prep.). Here we present only a short outline of the concept. The timeline $I(t)$ has reduced dimensionality compared to the $I(x,y)$ position-position space, but the noise spectrum is more complex.

\begin{equation}
I(t) = N_{(t)}^{1/f} + N_{(t)}^{\textrm{D}} + N_{(t)}^\textrm{det} + I_{(t)}^S
\label{eq1}
\end{equation}

In Eq. (\ref{eq1}) $N_{(t)}^{1/f}$ stands for the $1/f$ noise, $N_{(t)}^\textrm{D}$ is the drift noise, $N_{(t)}^\textrm{det}$ is the detector noise and the source intensity is expressed by $I_{(t)}^S$. In the L1 timeline we also have a lower $S/N$ than what we have in the L2 reconstructed maps. We are mainly interested in the subtraction of the high-frequency components. The L2 source masks are back-projected to the L1 timeline based on the celestial coordinates assigned to each L1 readout. The mask sizes are chosen to fully cover the apparent extent of a given source. We use stationary wavelet transform to decompose the signal, and to separate the low frequency baseline and the high frequency noise. This high frequency component is then used to Monte-Carlo simulate the noise on the masked section of the timeline. The algorithm applies a Monte-Carlo simulated (representative $1/f$) noise for the high-frequency part of the interpolated section. The reason behind is that the timeline noise properties would be discontinued by a simple interpolation (e.g. applying a polynomial or a smooth baseline). Map reconstruction techniques may change their response in terms of introducing local artefacts (aka reconstruction noise) around the masked area, and change noise properties all over the image.

\begin{figure}[!htbp]
\centering
  \includegraphics[width=8cm,angle=270]{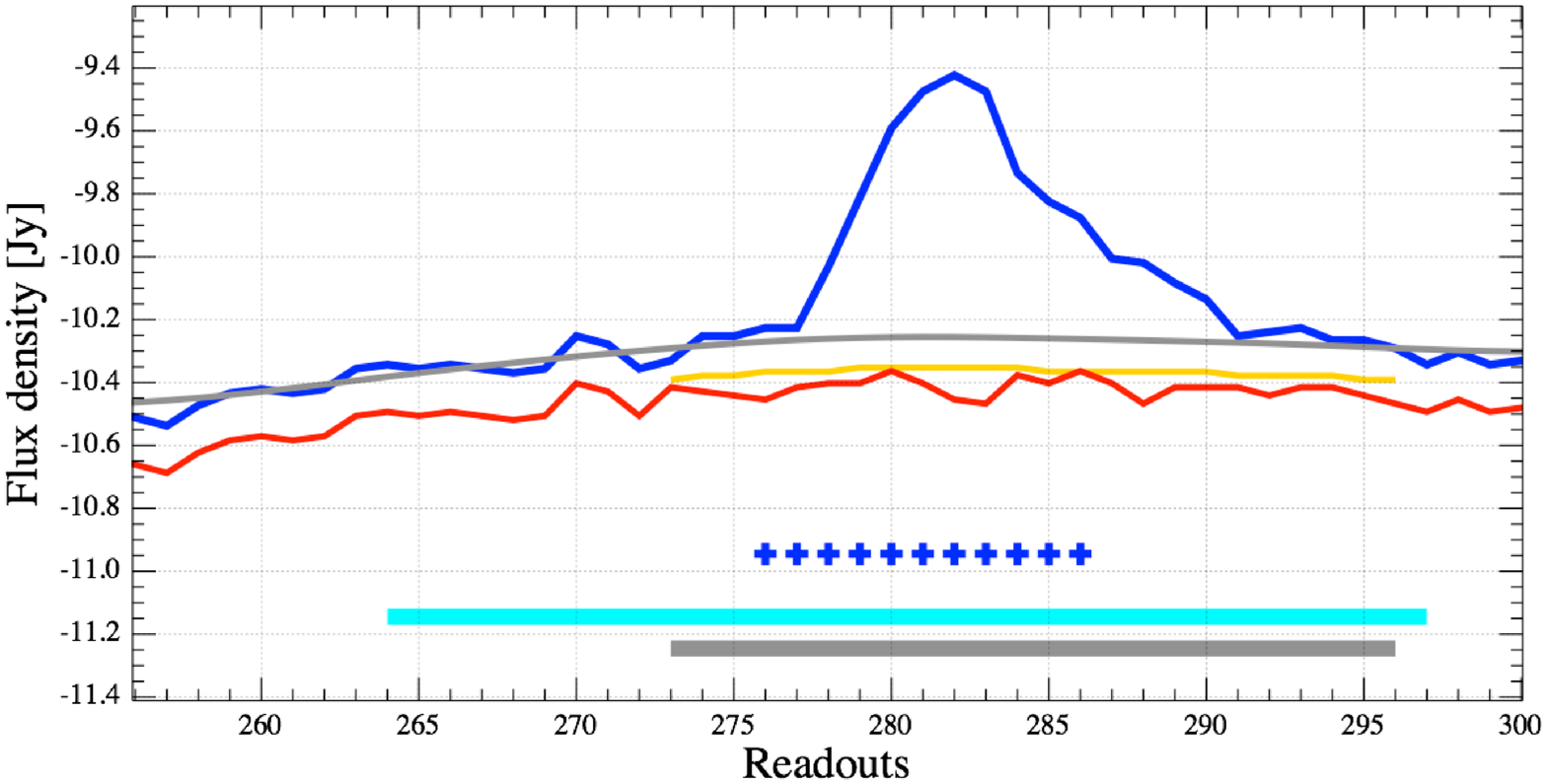}
\caption{A part of the L1 timeline with a contribution of a compact source in the middle. Blue solid line represent the original timeline. Gray solid line represent the estimated baseline (sky background), red and yellow lines represent the Monte-Carlo simulated and Gaussian noise shifted with $-0.1$ and $-0.2$ mJy for better visibility, respectively. Blue crosses, cyan and gray thick lines at the bottom of the figure represent the back-projected input mask defined by the user, the extended intermediate working mask and the optimized mask, respectively.}
\label{fig:1}       
\end{figure}

The size of the masked section is adaptive. The initial mask is first extended by a user-given factor and the determination of the final size is based on the first and last readouts which are below the estimated baseline. The interpolated signal on this section of the timeline is then produced as the sum of the low frequency baseline (sky background) and the Monte-Carlo simulated noise. An example of the timeline interpolation procedure is shown in Fig. \ref{fig:1}. This technique also allows us to separate the source intensity from the extended emission and to create source-only and source-free maps at the same time. 

\subsection{Comparison method}
\begin{figure}[!htbp]
\centering
  \includegraphics[width=4.5cm, angle=270,trim=3cm 0cm 6cm 0cm, clip=true]{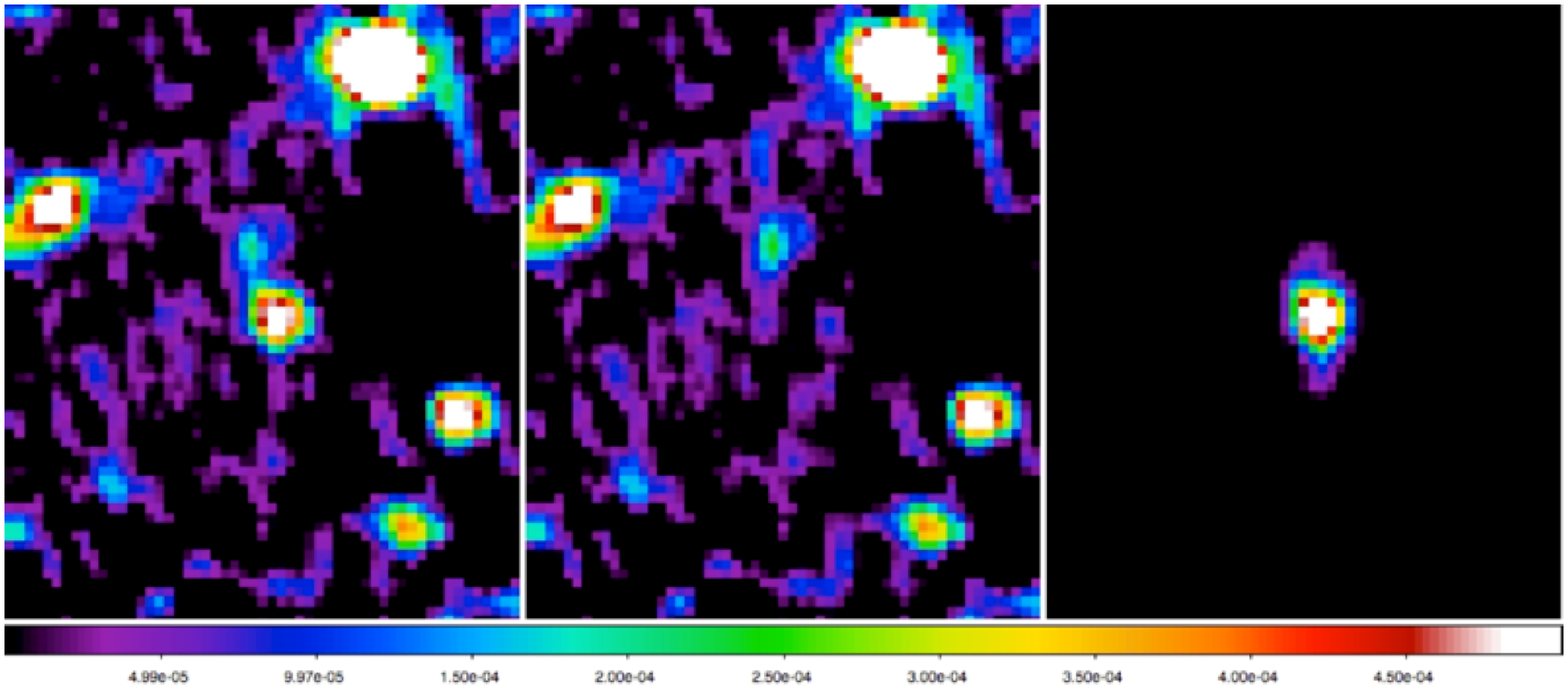}
\caption{Map of HD 170693 observed at 160 $\mu$m. The three panels are maps created using the original timeline (left), the interpolated timeline (middle) and the source-only timeline (right).}
\label{fig:2}       
\end{figure}

To find out if the algorithm is working properly, one needs to run various tests. As a first test, we subtracted standard stars with known brightness values from the background, followed by the re-projection of the timeline that contains only the source flux. These sources were measured by the PACS instrument \cite{poglitsch} in mini-scanmap mode \cite{msmm}, with $20^{\prime\prime}/$s scan-speed. More information on the mini scan-map mode and on the absolute flux calibration can be found in \cite{bz}, while details about the general scan-map technique can be found in \cite{gmmt}. The reason of these observations was to test the PACS photometry on faint, but reliable stars. Their predicted brightness at the three nominal PACS wavelengths were calculated from photospheric models. The values are listed in Table \ref{tab:1}. Note that the blue detector array observed either in the blue band (70$\mu$m) or in the green band (100 $\mu$m), in the case of a specific OBSID, while the red data was always taken. It means that we have a mixed sample of blue and green band maps for the short wavelength detector array. As an example, we present an original, a source-free and a source-only map in Fig. \ref{fig:2}. 

\begin{figure}[!htbp]
\centering
  \includegraphics[width=9cm,angle=90]{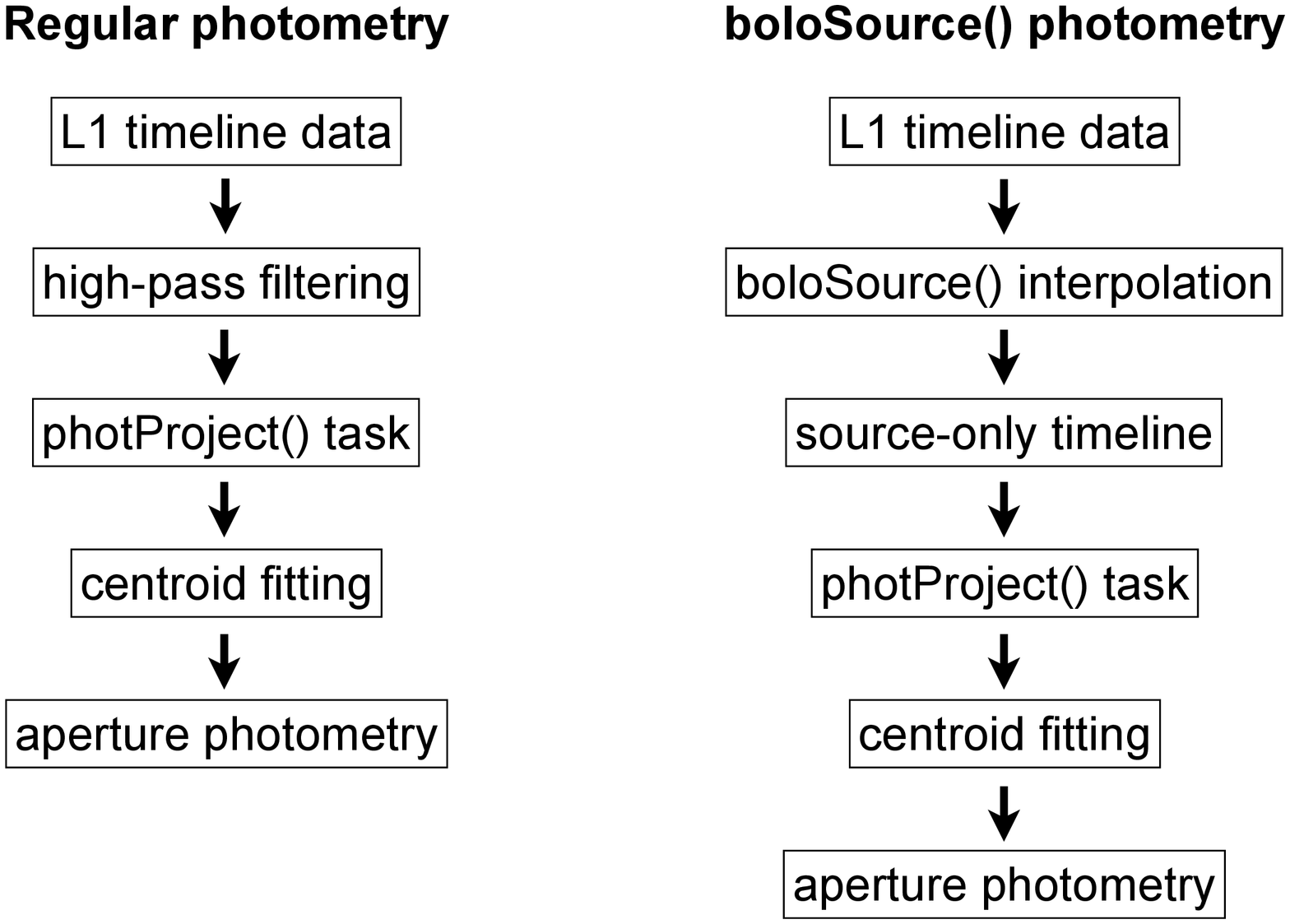}
\caption{Flowchart of the regular photometry produced on maps with nominal sky including background contamination (left chain) and the flowchart of the boloSource() photometry on the background-free maps (right chain). For details see text.}
\label{photflow}       
\end{figure}

In the first step the reliability of the standard star brightness was checked with the following method. Each observation consists of multiple repetitions. Starting from the first one, 2 consecutive repetitions were merged then high-pass filtering was used with high-pass filter width of 16, 18 and 36 readouts and were projected with the photProject task using pixel sizes of $1.1^{\prime\prime}$, $1.4^{\prime\prime}$ and $2.1^{\prime\prime}$ for the 70, 100 and 160 $\mu$m observations, respectively. The observed flux was then calculated with the built-in aperture photometry pipeline of the HIPE \cite{ott} and also with the use of the IDL (Interactive Data Language) "aper" routine. While the HIPE task includes aperture correction, for the IDL photometry we used the encircled energy fraction (EEF) values listed in Table 15 of \cite{eef}. We also applied color correction assuming a black body temperature of 5000 K. For the flowchart of the regular photometry see the left column of Fig \ref{photflow}. From the measured points a light curve was constructed in each case and the standard deviation ($\sigma$) of the light curve points was calculated. In order to rule out the uncertainty of the instrument pointing, we constructed light curves with different aperture radius sizes ranging from $2^{\prime\prime}$ to $10^{\prime\prime}$, and checked if the photometric stability increases with aperture size. The average brightness of a given light curve was also calculated and was compared to the predicted one. The light curves are characterized by the standard deviation of the point source fluxes around the average ($\sigma$) and by the correction factor $C$ which is the ratio of the average measured flux to the predicted flux ($C=\overline{F}_\textrm{meas}/F_\textrm{pred}$).


In the second step the source-only maps were created with the photProject task without the application of the high-pass filter method, with the same pixel sizes as given earlier. We determined the boloSource() flux values with the IDL "aper" routine and described the light curves in the same way as we did for the untouched maps. For the flowchart of the boloSource() photometry see the right column of Fig \ref{photflow}. For consistency and comparison we used the brightness values derived with IDL "aper" routine on both the original and source-only maps.

%
\begin{table}[!htbp]
\begin{center}
\caption{Parameters of the standard stars used in testing the photometric capabilities of the boloSource() algorithm. Columns are as follows: source ID, spectral type and predicted brightness at 70, 100 and 160 $\mu$m with their corresponding uncertainty, respectively. Predicted flux values and their uncertainties at 70 $\mu$m are taken from \cite{gordon}. Values for the 100 and 160 $\mu$m were calculated. Uncertainties for HD 139669 were not available.}
\label{tab:1}       
\begin{tabular}{|c|c|ccc|}
\hline
Name & SpType & 70$\mu$m [mJy] & 100$\mu$m [mJy] & 160$\mu$m [mJy]  \\
\hline
HD 15008 & A1/2V & 22.0 $\pm$ 0.8 & 10.8 $\pm$ 0.4 & 4.2 $\pm$ 0.2 \\
HD 152222 & K2III&37.9 $\pm$ 1.8 & 18.6 $\pm$ 0.9 & 7.3 $\pm$  0.3\\
HD 39608 & K5III &29.6 $\pm$ 1.2 & 14.6 $\pm$ 0.6 & 5.7 $\pm$ 0.2\\
HD 159330 & K2III & 61.7 $\pm$ 2.0 & 30.2 $\pm$ 1.0 & 11.8 $\pm$ 0.4\\
HD 139669 & K5III & 286 & 140 & 54 \\
HD 138265 & K5III &111.4 $\pm$ 3.8 & 54.6 $\pm$ 1.9 & 21.3 $\pm$ 0.7\\
HD 170693 & K1.5III &147.7 $\pm$ 4.4 & 72.4 $\pm$ 2.2 & 28.3 $\pm$ 0.8\\
\hline
\end{tabular}
\end{center}
\end{table}

\section{Results}
\label{sec:2}

\begin{figure}
\centering
  \includegraphics[width=10cm]{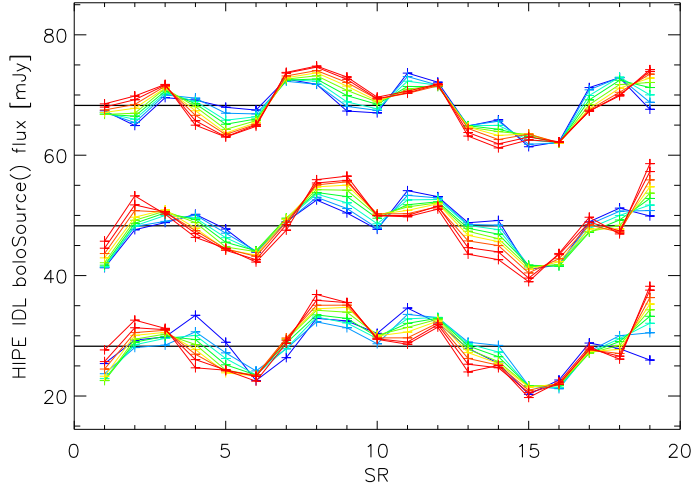}
\caption{HIPE (bottom), IDL (middle) and boloSource() (top) light curves of HD 170693 at 160$\mu$m as a function of repetition number. Colours from blue to red indicate the different aperture radii from $2^{\prime\prime}$ to $10^{\prime\prime}$. IDL and boloSource() values are shifted with +20 and +40 mJy for better visibility.}
\label{fig:3}       
\end{figure}

As the example light curves show in Fig. \ref{fig:3}, the brightness of the standard stars show some variation along the repetitions. This variation depends neither on the method used to derive the photometric values, nor on the size of the aperture radius. Therefore, we did not calculate all the brightness values with all aperture radii, but only with $5.0^{\prime\prime}$, $6.0^{\prime\prime}$ and $7.0^{\prime\prime}$  at 70, 100 and 160$\mu$m, respectively. Comparison of the different methods were made in each band, individually. 

\begin{figure}
\centering
\begin{tabular}{cc}
  \includegraphics[width=0.5\textwidth]{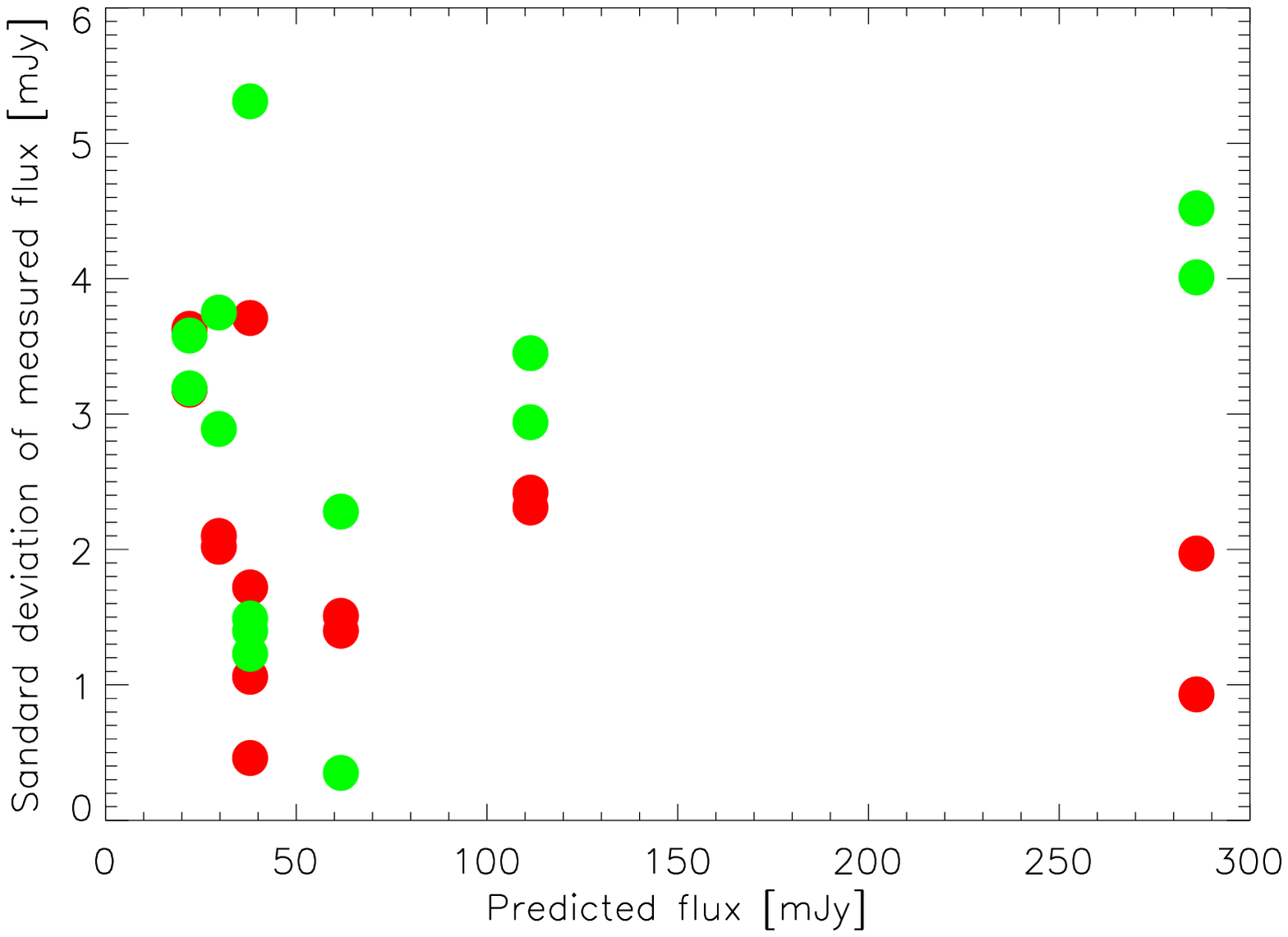}
  \includegraphics[width=0.5\textwidth]{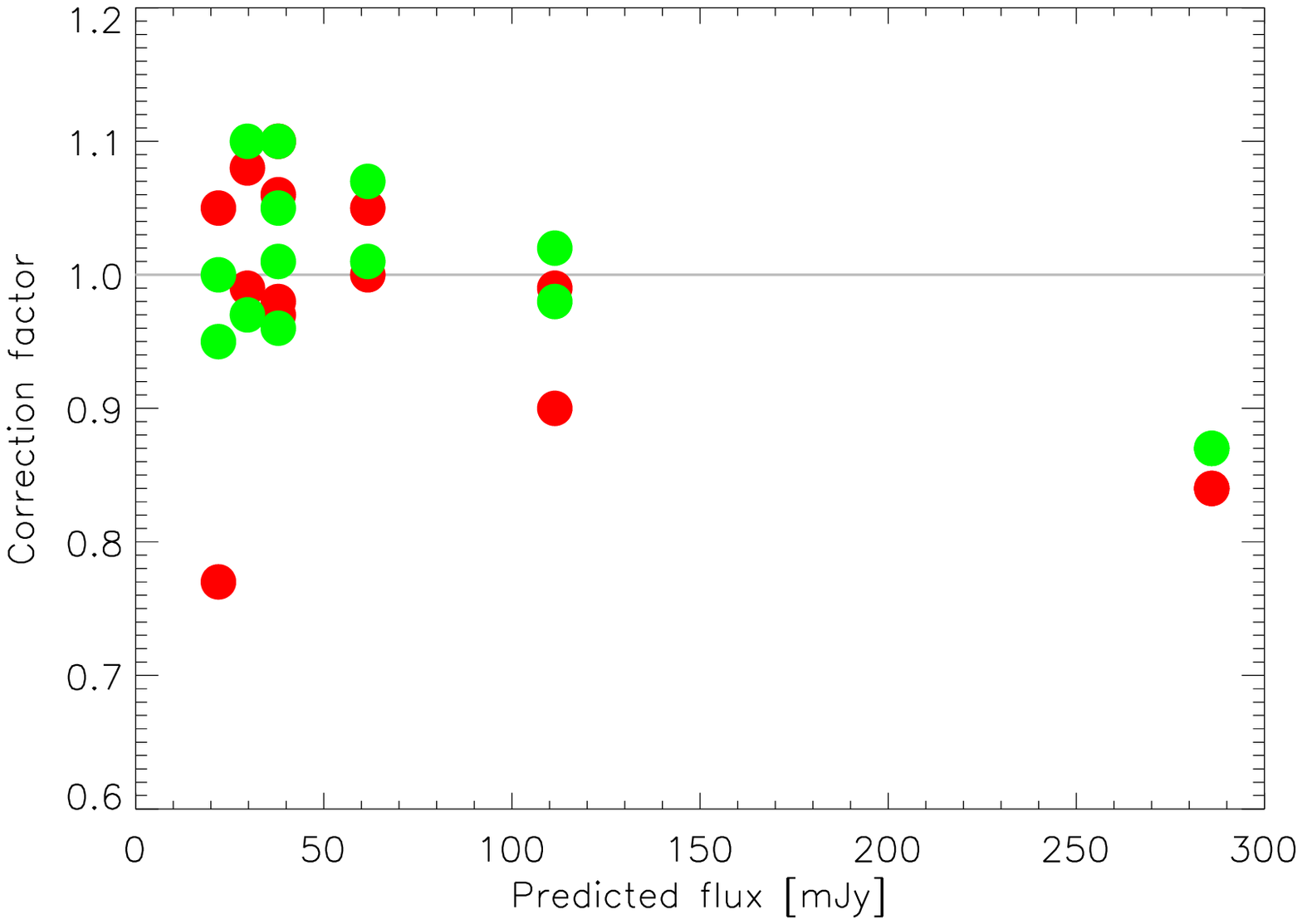}
 \end{tabular}
\caption{$\sigma$ (left panel) and $C$ (right panel) values plotted as a function of the predicted brightness in the 70$\mu$m case. Red filled circles present the values calculated from the original maps, green symbols present the boloSource() values.}
\label{fig:4}       
\end{figure}
\begin{figure}
\centering
\begin{tabular}{cc}
  \includegraphics[width=0.5\textwidth]{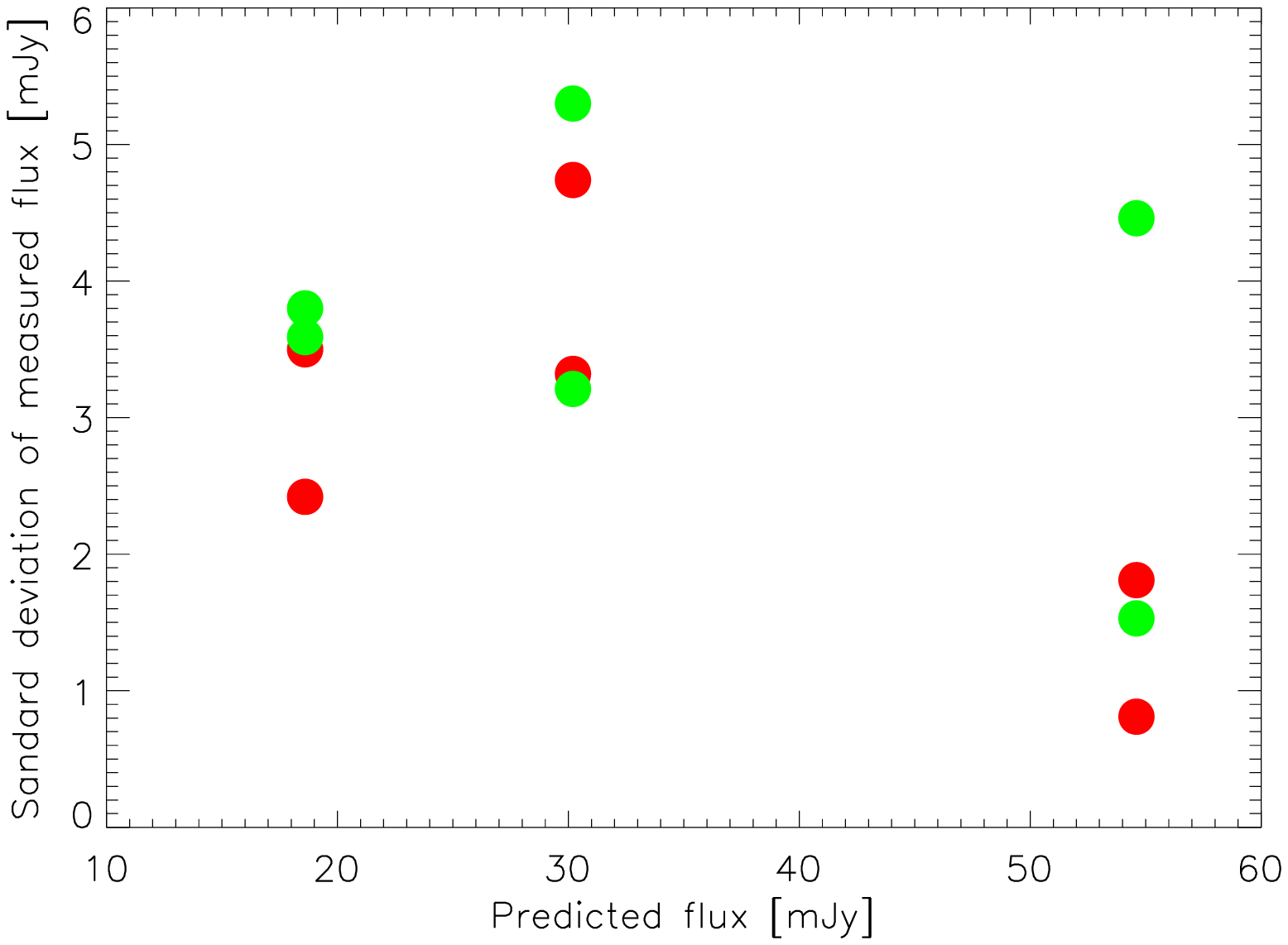}
  \includegraphics[width=0.5\textwidth]{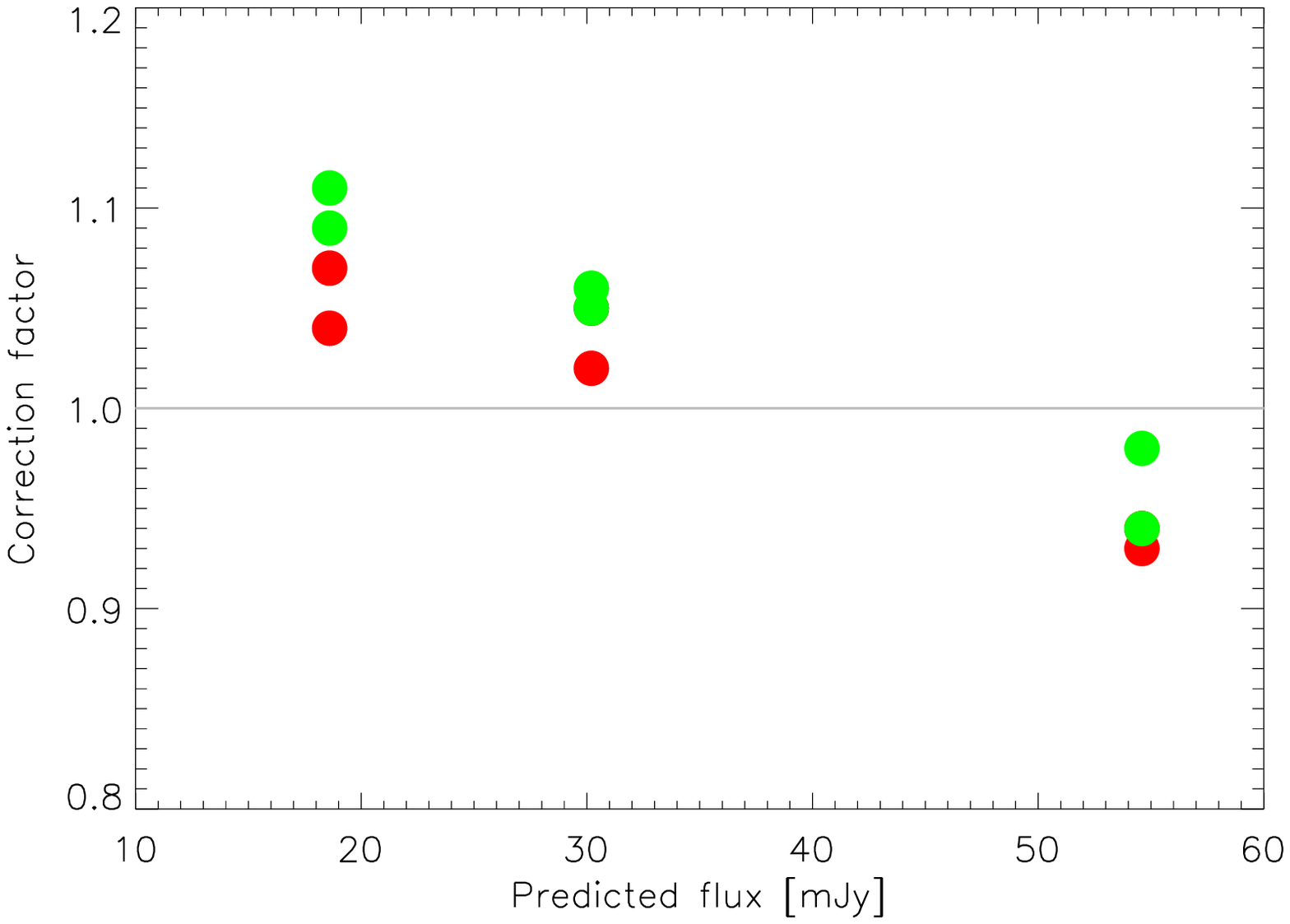}
 \end{tabular}
\caption{Same as Fig. \ref{fig:4}, for the 100$\mu$m case.}
\label{fig:5}       
\end{figure}
\begin{figure}
\centering
\begin{tabular}{cc}
  \includegraphics[width=0.5\textwidth]{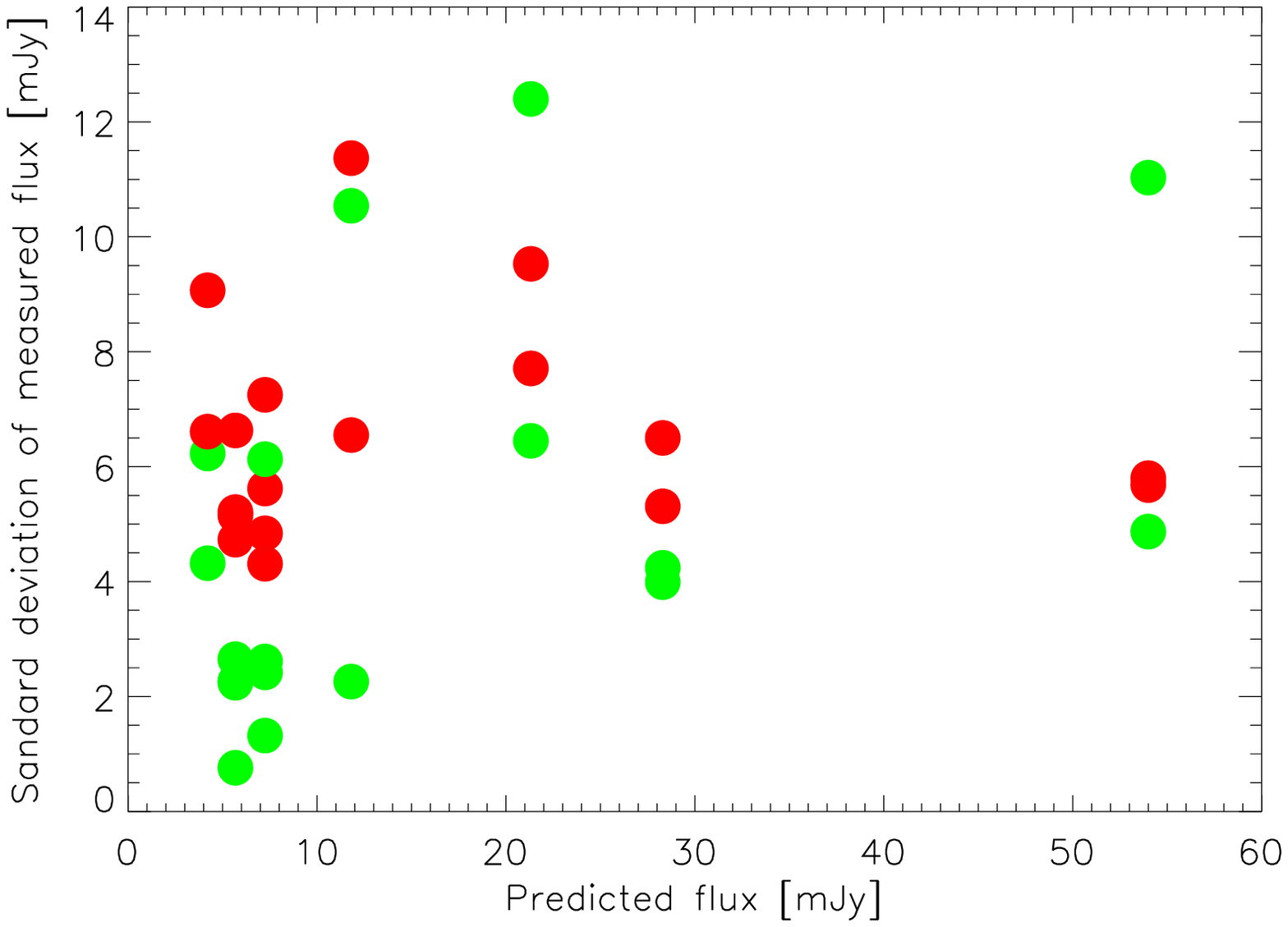}
  \includegraphics[width=0.5\textwidth]{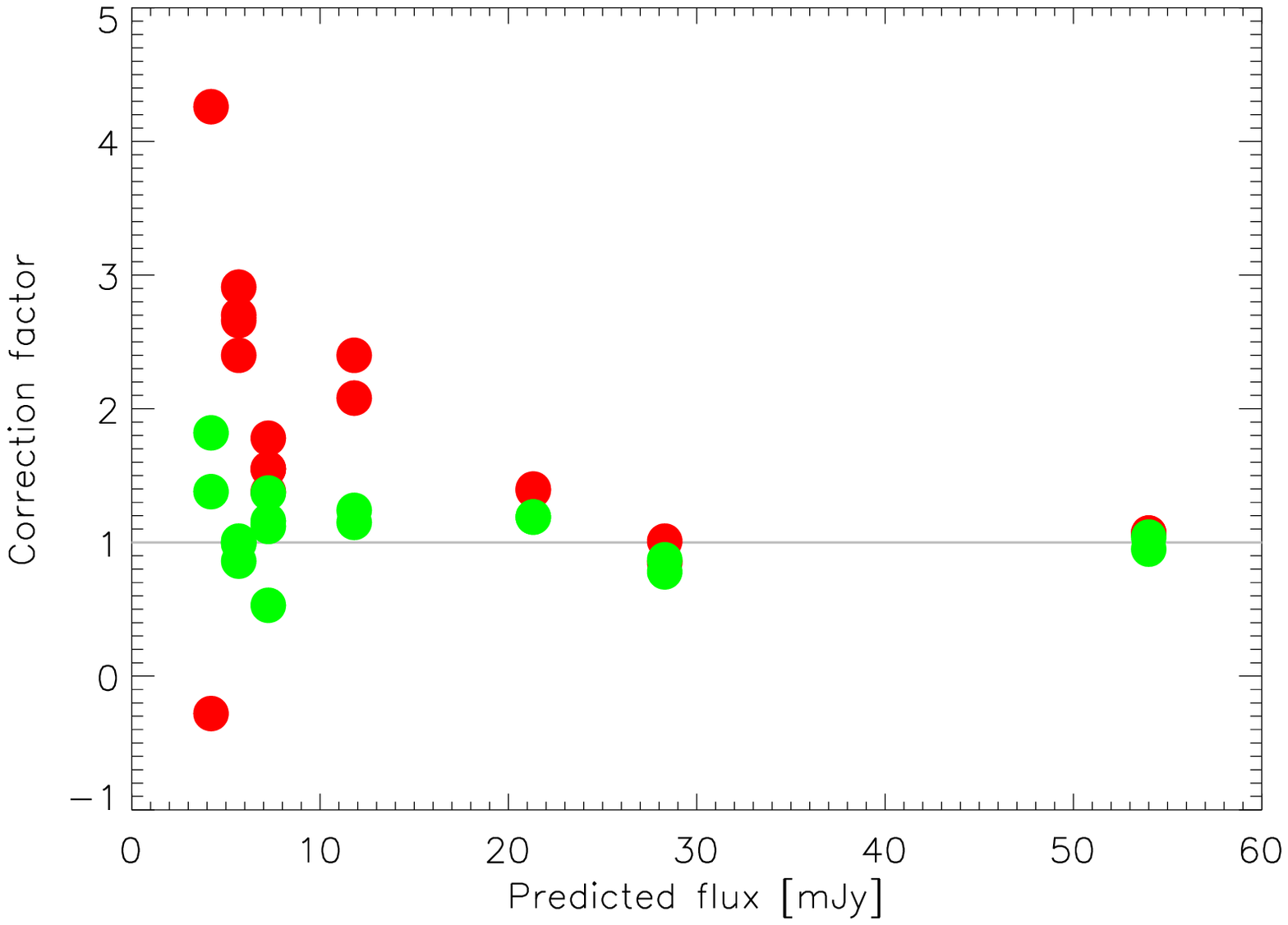}
 \end{tabular}
\caption{Same as Fig. \ref{fig:4}, for the 160$\mu$m case.}
\label{fig:6}       
\end{figure}

\begin{itemize}
\item The results obtained at 70$\mu$m are listed in Table \ref{tab:2} and are plotted in Fig. \ref{fig:4}. The median values of $\sigma$ (the standard deviations of the point source flux) were found to be 2.02 mJy in the case of the original maps and 3.19 mJy in the boloSource() case, while the averages were found to be 2.03 mJy and 2.89 mJy. The median of $C$ correction factors are 0.99 and 1.01 in the original and boloSource() case, respectively, while the averages are 0.97 and 1.00. 

\item 100$\mu$m results are listed in Table \ref{tab:3} and are plotted in Fig. \ref{fig:5}. For the original and for the boloSource() cases the median $\sigma$ values were found to be 3.32 mJy and 3.80 mJy, while the average $\sigma$ values were found to be 2.77 mJy and 3.65 mJy. The median $C$ values are 1.04 and 1.06, while the average $C$ values are 1.01 and 1.04. 

\item The 160$\mu$m results are listed in Table \ref{tab:4} and are plotted in Fig. \ref{fig:6}. The median $\sigma$ values from original and boloSource() cases are 6.50 mJy and 4.24 mJy, while the average $\sigma$ values were found to be 6.54 mJy and 4.81 mJy. The median $C$ values are 1.55 and 1.12, while the averages are 1.78 and 1.09. 
\end{itemize}

%

\begin{table}[!htbp]
\begin{center}
\caption{$\sigma$ and $C$ parameters calculated for the standard star light curves at 70$\mu$m. Columns are as follows: 1)Standard star name, 2)Observational ID of the Herschel observation from which the light curve was derived, 3)Number of repetitions in the measurement, 4)Predicted brightness of the star, 5)-6)$\sigma$ and $C$ values calculated for IDL brightness values, 7)-8)$\sigma$ and $C$ values calculated for boloSource() brightness values.}
\label{tab:2}       
\small\addtolength{\tabcolsep}{-3pt}
\begin{tabular}{|c|c|c|c|c|c|c|c|}
\hline
Name & OBSID & Repetitions & Brightness [mJy]& $\sigma_\textrm{IDL}$ & $C_\textrm{IDL}$ & $\sigma_\textrm{bS()}$	& $C_\textrm{bS()}$ \\
\hline
HD 15008	& 1342189130	& 9 & 22.0 & 3.63 & 1.05 & 3.58 & 1.00\\ 
HD 15008 & 1342189131 & 9 & 22.0 & 3.18 & 0.77 & 3.19 & 0.95\\ 
HD 152222 & 1342240702 & 6 & 37.9 & 3.71 & 0.98 & 5.31 & 1.01\\
HD 152222 & 1342240703 & 6 & 37.9 & 1.72 & 1.10 & 1.40 & 1.10\\
HD 152222 & 1342191964 & 6 & 37.9 & 1.06 & 0.97 & 1.23 & 0.96\\
HD 152222 & 1342191965 & 6 & 37.9 & 0.46 & 1.06 & 1.49 & 1.05\\
HD 39608 & 1342198535 & 10 & 29.7 & 2.10 & 1.08 & 3.75 & 1.10\\
HD 39608 & 1342198536 & 10 & 29.7 & 2.02 & 0.99 & 2.89 & 0.97\\
HD 159330 & 1342213585 & 6 & 61.7 & 1.40 & 1.00 & 0.35 & 1.01\\
HD 159330 & 1342213586 & 6 & 61.7 & 1.51 & 1.05 & 2.28 & 1.07\\
HD 139669 & 1342191982 & 6 & 286 & 0.93 & 0.84 & 4.01 & 0.87\\
HD 139669 & 1342191983 & 6 & 286 & 1.97 & 0.84 & 4.52 & 0.87\\
HD 138265 & 1342188841 & 21 & 111.4 & 2.31 & 0.90 & 2.94 & 0.98\\
HD 138265 & 1342188842 & 21 & 111.4 & 2.42 & 0.99 & 3.45 & 1.02\\
\hline
\end{tabular}
\end{center}
\end{table}

\begin{table}[!htbp]
\begin{center}
\caption{Same as Table \ref{tab:2}, for the 100$\mu$m case.}
\label{tab:3}       
\small\addtolength{\tabcolsep}{-3pt}
\begin{tabular}{|c|c|c|c|c|c|c|c|}
\hline
Name & OBSID & Repetitions & Brightness [mJy]& $\sigma_\textrm{IDL}$ & $C_\textrm{IDL}$ & $\sigma_\textrm{bS()}$	& $C_\textrm{bS()}$ \\
\hline
HD 152222 & 1342227973 & 33 & 18.6 & 2.42 & 1.04 & 3.59 & 1.09\\
HD 152222 & 1342227974 & 33 & 18.6 & 3.50 & 1.07 & 3.80 & 1.11\\
HD 159330 & 1342188839 & 10 & 30.2 & 4.74 & 1.02 & 5.30 & 1.05\\
HD 159330 & 1342188840 & 10 & 30.2 & 3.32 & 1.05 & 3.21 & 1.06\\
HD 138265 & 1342191986 & 6 & 54.6 & 1.81 & 0.94 & 1.53 & 0.98\\
HD 138265 & 1342191987 & 6 & 54.6 & 0.81 & 0.93 & 4.46 & 0.94\\
\hline
\end{tabular}
\end{center}
\end{table}

\begin{table}[!htbp]
\begin{center}
\caption{Same as Table \ref{tab:2}, for the 160$\mu$m case.}
\label{tab:4}       
\small\addtolength{\tabcolsep}{-3pt}
\begin{tabular}{|c|c|c|c|c|c|c|c|}
\hline
Name & OBSID & Repetitions & Brightness [mJy]& $\sigma_\textrm{IDL}$ & $C_\textrm{IDL}$ & $\sigma_\textrm{bS()}$	& $C_\textrm{bS()}$ \\
\hline
HD 15008 & 1342189130 & 9 & 4.21 & 6.58 & -0.05 & 7.78 & 1.07\\ 
HD 15008 & 1342189131 & 9 & 4.21 & 7.57 & 0.93 & 6.20 & 1.00\\ 
HD 152222 & 1342240702 & 6 & 7.25 & 0.05 & 0.90 & 2.21 & 0.67\\ 
HD 152222 & 1342240703 & 6 & 7.25 & 5.79 & 1.37 & 3.80 & 1.38\\ 
HD 152222 & 1342191964 & 6 & 7.25 & 3.98 & 1.26 & 1.58 & 1.30\\ 
HD 152222 & 1342191965 & 6 & 7.25 & 4.49 & 1.41 & 2.71 & 1.26\\ 
HD 39608 & 1342198535 & 10 & 5.68 & 4.21 & 1.87 & 0.82 & 1.28\\ 
HD 39608 & 1342198536 & 10 & 5.68 & 5.21 & 1.90 & 2.33 & 1.15\\ 
HD 39608 & 1342198537 & 35 & 5.68 & 6.66 & 1.61 & 2.32 & 1.18\\
HD 39608 & 1342198538 & 35 & 5.68 & 5.14 & 1.99 & 2.76 & 1.05\\
HD 159330 & 1342213585 & 6 & 11.81 & 3.17 & 1.26 & 1.97 & 1.09\\ 
HD 159330 & 1342213586 & 6 & 11.81 & 4.04 & 0.72 & 7.57 & 0.91\\ 
HD 139669 & 1342191982 & 6 & 54.0 & 5.31 & 1.01 & 7.72 & 1.00\\ 
HD 139669 & 1342191983 & 6 & 54.0 & 7.57 & 0.97 & 7.54 & 0.89\\ 
HD 138265 & 1342188841 & 21 & 21.32 & 8.22 & 0.83 & 9.93 & 1.11\\
HD 138265 & 1342188842 & 21 & 21.32 & 7.38 & 0.81 & 4.79 & 1.11\\
HD 170693 & 1342243730 & 20 & 28.3 & 6.50 & 0.85 & 3.99 & 0.78\\
HD 170693 & 1342243731 & 20 & 28.3 & 5.31 & 1.01 & 4.24 & 0.87\\
\hline
\end{tabular}
\end{center}
\end{table}

\section{Summary}
In this work we demonstrated the use of the boloSource() on low and flat background. Point sources were subtracted and source-only maps were created from timeline data that contained only the source flux. Comparison of photometry on the source-only maps and on the original maps was made. This analysis is a first step in our testing of the boloSource() capabilities, and we were able to draw statistically reliable uncertainties. Such high repetition factor observations do not exist for crowded fields. Therefore, well isolated single point source measurements with multiple repetitions were used. 

We conclude that on low and flat background the results of our method are fully comparable to the regular aperture photometry, the average and the median of $\sigma$ values calculated with the two different methods agree well within $\sim$1 mJy difference. Given the $\sim$5\% absolute flux accuracy of the measurements \cite{eef} and the uncertainty of the predicted brightness values, the comparison of the $C$ correction factors show that boloSource() is also reliable in recovering the predicted absolute flux. From shorter to longer wavelengths the flux uncertainty is decreased using our method compared with the regular photometry, while the accuracy of the recovered flux is increased. 

We note that boloSource() is still under development with the goal of making it usable for large-scale studies. The test environment need to be extended to simulated data sets that can mimic the low-redundancy complex galactic fields. We plan to add simulated point sources in a wide flux range to a simulated cirrus background. Test cases will cover various positions on the surface-brightness vs. background complexity parameter space. We expect that boloSource() provides an increasing efficiency towards the more complex cases (better than aperture photometry) but this statement is to be justified. Our goals also include an extension to the SPIRE photometer observations. Detailed description of the algorithm, and the further tests and results are subject of a future paper.

\begin{acknowledgements}
We thank our anonymous referee whose comments significantly improved the manuscript. We thank Z. Balog for the useful discussion. This work has been supported by the following grants: i) PECS contract no. 98073 of the Hungarian Space Office and the European Space Agency, ii) Hungarian Research Fund (OTKA) grants nr. 101939 and 104607. C.K. acknowledges the support of the Bolyai Research Fellowship of the Hungarian Academy of Sciences. We acknowledge support from the Faculty of the European Space Astronomy Centre (ESAC).
\end{acknowledgements}




\end{document}